\documentclass[aps,a4paper,amsmath,amssymb,twocolumn,
floatfix,groupedaddress]{revtex4}
\usepackage{bm,amsmath,amssymb,amsfonts}
\usepackage[dvips]{graphicx}
\usepackage{color}
\definecolor{dblue}{rgb}{0.0,0.0,0.7}
\definecolor{dred}{rgb}{0.9,0.0,0.0}
\usepackage{hyperref}

\hypersetup{
    pdfnewwindow=true,      % links in new window
    colorlinks=true,       % false: boxed links; true: colored links
    linkcolor=dred,          % color of internal links
    citecolor=blue,        % color of links to bibliography
    filecolor=blue,      % color of file links
    urlcolor=blue        % color of external links
}
\newcommand\bea{\begin{eqnarray}}
\newcommand\eea{\end{eqnarray}}
\newcommand\beq{\begin{equation}}
\newcommand\eeq{\end{equation}}

\newcommand{\ie}{{\it i.\,e., }}

\newcommand{\etal}{{\it et al.}}
\newcommand{\viz}{{\it viz., }} 

\begin{document}

% Title 1
\title{Flat bands in fractal-like geometry }

\author{Biplab Pal} 
\email{biplabpal@pks.mpg.de}
\author{Kush Saha}
\email{kush@pks.mpg.de}
\affiliation{Max Planck Institute for the Physics of Complex Systems, N\"{o}thnitzer Str.\ 38, 
01187 Dresden, Germany }

\begin{abstract}
We report the presence of multiple flat bands in a class of two-dimensional (2D) lattices 
formed by Sierpinski gasket (SPG) fractal geometries as the basic unit cells. Solving the 
tight-binding Hamiltonian for such lattices with different generations of a SPG network, 
we find multiple degenerate and non-degenerate {\it completely flat} bands, depending on 
the configuration of parameters of the Hamiltonian. Moreover, we establish a generic formula 
to determine the number of such bands as a function of generation index, $\ell$ of the 
fractal geometry. We show that the flat bands and their neighboring dispersive bands have 
remarkable features, the most interesting one being the spin-1 conical-type spectrum at the 
band center without any staggered magnetic flux, in contrast to the Kagome lattice. 
We furthermore investigate the effect of magnetic flux in these lattice settings 
and show that different combinations of fluxes through such fractal unit cells lead to 
richer spectrum with a single isolated flat band or gapless electron- or hole-like flat 
bands. Finally, we discuss a possible experimental setup to engineer such fractal flat 
band network using single-mode laser-induced photonic waveguides.  
\end{abstract}

% \pacs{71.20.-b, 63.20.Pw, 64.60.al}
\maketitle

%%%%%%%%%%%%%%%%%%%%%%%%%%%%%%%%%%%%%%%%%%%%%%%%%%%%%%%%%%%%%%%%%%%%%%%%%%%%%%%%%%%%%%%%%%%%%%%%%%%%%%%%%%%%%%%%%%%%
\section{Introduction}
Over the course of last few years, the study of flat bands (FBs) in translationally 
invariant lattice systems has been one of the emergent fields of research in condensed 
matter physics~\cite{goda-prl2006, shukla-prb2010, goda-jpsj2007, goda-jpsj2003, 
goda-jpsj2005, manninen-pra2013, takashi-prb2010, aldea-prb2013, flach-prb2013, flach-epl2014, 
flach-prl2014, flach-prb2015, flach-prl2016,flach-prb2017,ajith-prb2017}. This is in 
part because such bands can serve as a good platform for studying strongly correlated 
phenomena due to exponentially large number of degeneracy, and in part because they 
can host many interesting phenomena such as unconventional inverse Anderson 
transition~\cite{goda-prl2006, shukla-prb2010}, multifractality at weak disorder~\cite{goda-jpsj2007, 
aldea-prb2013}, Hall ferromagnetism~\cite{kimura-prb2002, tanaka-prl2003, pollman-prl2008}, etc.
In addition, there are rising interests in finding out lattice models that can host 
FBs with non-trivial topology. This is because such non-trivial FBs may allow 
one to investigate lattice version of fractional topological phenomena~\cite{kai-sun-prl2011, 
tang-prl2011, neupert-prl2011}. Above all, the recent experimental observations of FBs 
in various photonic lattices~\cite{vicencio-prl2015, mukherjee-prl2015, mukherjee-ol2015, 
longhi-ol2014, vicencio-njp2014, xia-ol2016, zong-oe2016, weimann-ol2016}, optical 
lattices~\cite{manninen-pra2010, bloch-rmp2008, goldman-pra2011}, and exciton-polariton 
condensates~\cite{masumoto-njp2012} have triggered a great deal of interest in generating 
new FB networks and understanding their usage in different lattice geometry.

The appearance of FBs in translationally invariant tight-binding models 
is often attributed to the destructive interference of electrons hopping nonlocally, 
giving rise to compact localized  single-particle eigenstates (CLS) with finite 
amplitudes over a finite number of lattice sites beyond which the wave function 
amplitudes decay to zero~\cite{flach-epl2014, flach-prl2016, flach-prb2017}. 
Specifically, the electrons in these FB states 
do not hop to the neighboring lattice sites, leading to strongly localized states
with an infinite effective band mass. FBs with non-trivial topology can be obtained in lattice models 
by fine-tuning the short-range hopping strengths between different lattice sites or by incorporating 
some artificial phase factors to the hopping parameters~\cite{kai-sun-prl2011, 
neupert-prl2011}. Recently, it has also been shown that FBs with chiral symmetry can be realized in bipartite 
lattices by creating imbalance in number of sites between the two sublattices~\cite{ajith-prb2017}.

%######################################################
\begin{figure}[ht]
\includegraphics[clip,width=0.75\columnwidth]{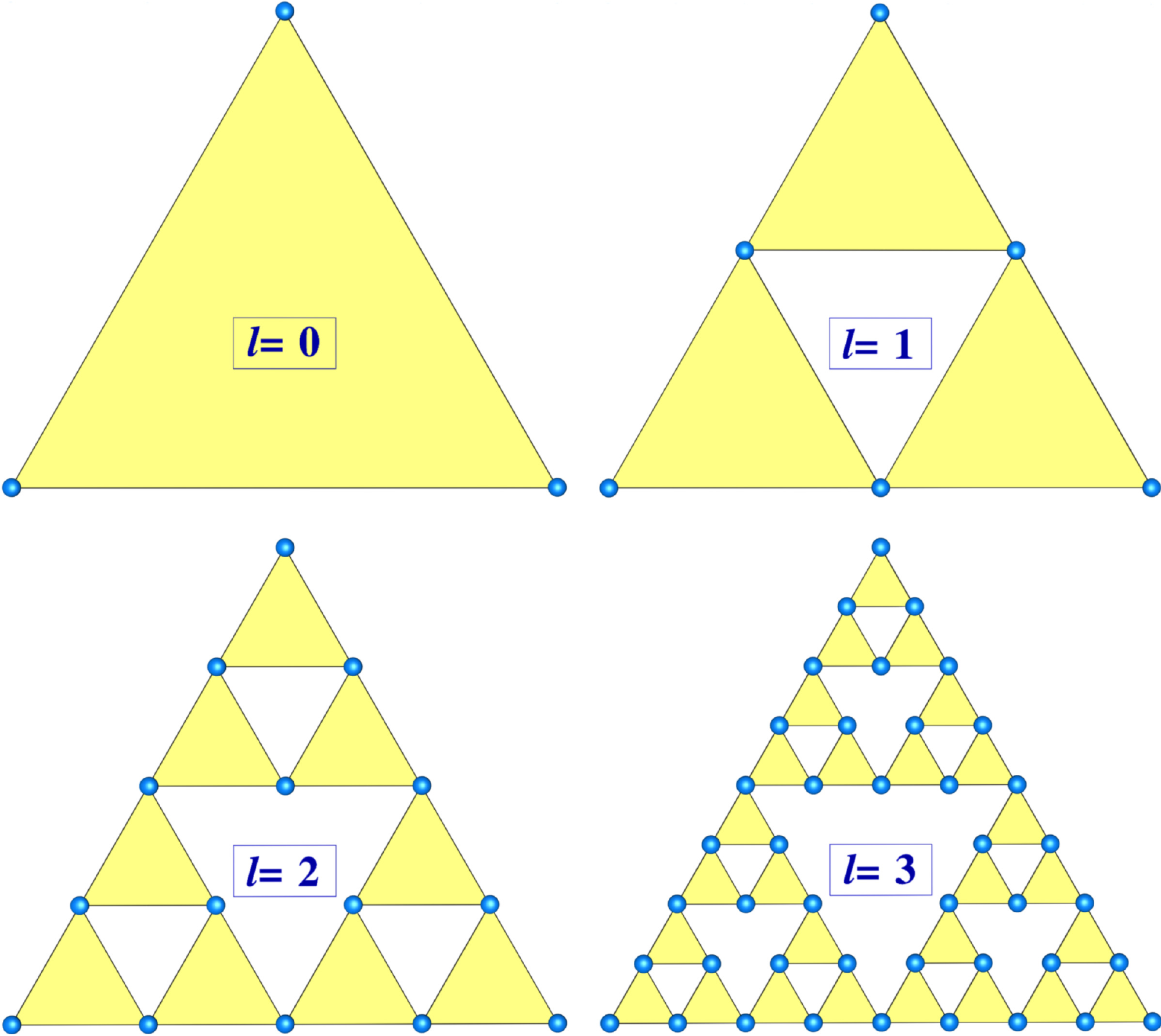}
\caption{Schematic view of different generations ($\ell$) 
of a Sierpinski gasket (SPG) fractal network, which are repeated 
periodically over a two-dimensional plane to form 2D lattices with 
fractal unit cell.}
\label{fig:fractal}
\end{figure}
%######################################################

%######################################################
\begin{figure}[ht]
\includegraphics[clip,width=\columnwidth]{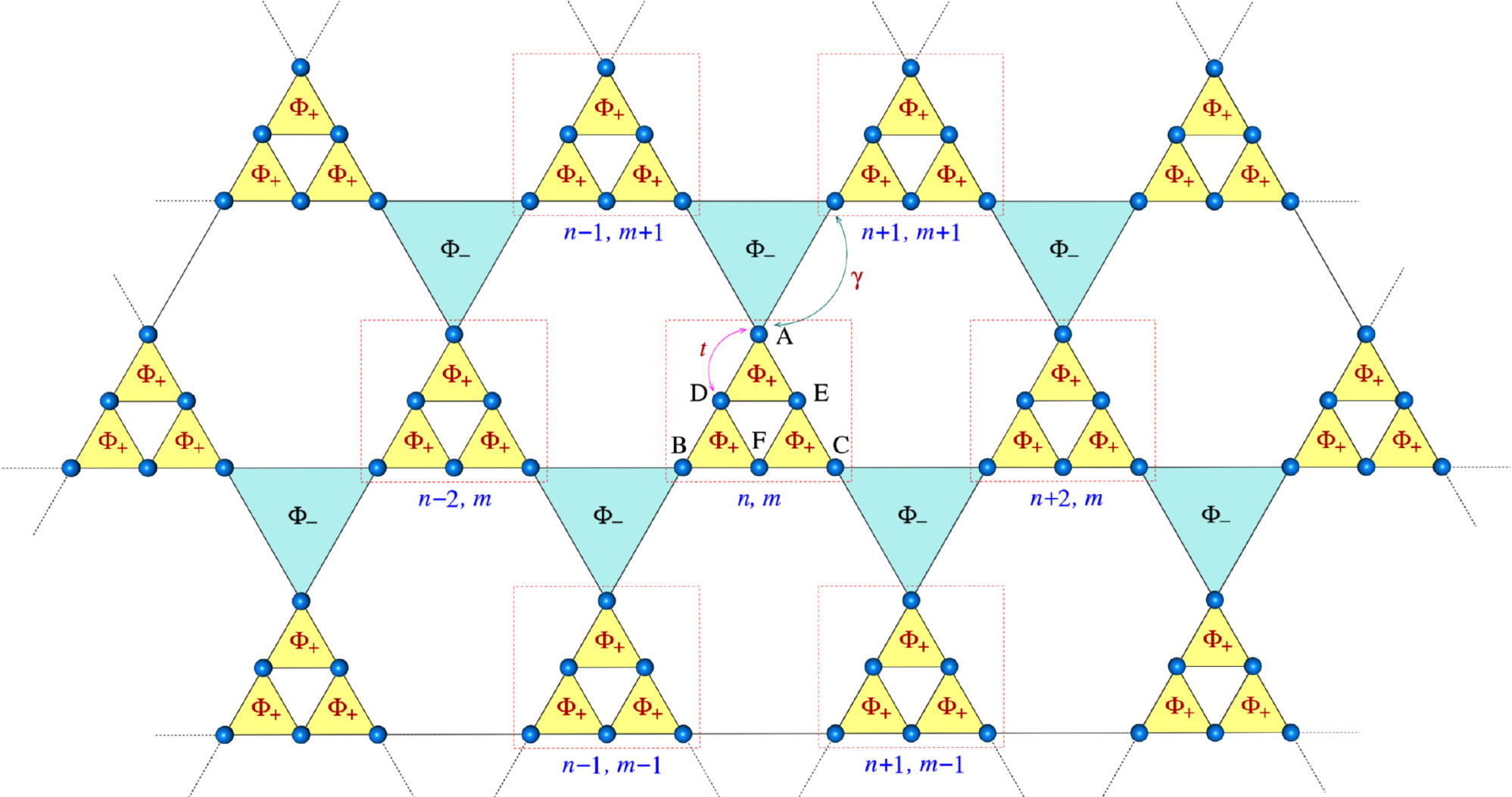}
\caption{Schematic diagram of a two-dimensional lattice structure where the 
building blocks (unit cells) are first generation ($\ell=1$) SPG fractal geometry. 
Each of the ``up" triangular plaquettes (light yellow shaded) within the SPG unit cell 
are threaded by an external magnetic flux $\Phi_{+}$, whereas the ``down'' triangles 
(light blue shaded) connecting nearest unit cells carry a flux $\Phi_{-}$. Each cell 
has been marked by the red dotted box. The intra-cell hopping parameter is 
denoted by $t$ and the inter-cell hopping parameter is labeled by $\gamma$.}
\label{fig:lattice}
\end{figure}
%######################################################

While there are extensive studies on FBs for regular periodic lattices such as 
Lieb~\cite{manninen-pra2010, aldea-prb2013, vicencio-prl2015, mukherjee-prl2015, 
vicencio-njp2014, goldman-pra2011}, Kagome~\cite{zong-oe2016, masumoto-njp2012, altman-prb2010, 
bergman-prb2008, green-prb2010, nagaosa-prb2000}, hexagonal ~\cite{goda-jpsj2003, 
wu-prl2007}, diamond ~\cite{flach-prb2013, flach-prl2016, mukherjee-ol2015, longhi-ol2014}, 
cross-switch~\cite{flach-epl2014, flach-prl2014, flach-prb2015}, etc., the possibility 
of having FBs with various interesting band features in two-dimensional (2D) lattices with fractal unit cells 
has not been reported yet. We show that the 2D lattices built with different 
generations of a Sierpinski gasket (SPG) fractal network~\cite{spg} (see Fig.~\ref{fig:fractal}) 
as the unit cells can host multiple degenerate and non-degenerate FBs, depending on 
the intra and inter-cell hopping strengths. These FBs and their nearest dispersive bands have distinct features such as 
``isolated" degenerate FBs, conical ``Dirac-like" bands together with a single flat band at the center of the energy 
spectrum, and many more. 
These band features differ significantly from the Kagome (\ie $\ell=0$ generation of the SPG) or Lieb lattice. 
We furthermore show that the number of such FBs increases with the increase of 
fractality \ie the generation index $\ell$ of the fractal unit cells. In fact, we 
find a generic formula to determine the number of FBs as a function of $\ell$. 

We also investigate the effect of an external magnetic field in this lattice model and 
show that the different combinations of magnetic fluxes piercing through the ``up" 
and ``down" triangles of the lattice structure (see Fig.~\ref{fig:lattice}) give 
rise to spectrum with notable features. They are classified into three types: 
(\rm I) a {\it single} isolated flat band at the middle of the band spectrum, 
separating electron-like and hole-like dispersive bands, (\rm II) a {\it single 
gapless} hole-like or electron-like flat band, and (\rm III) two {\it gapless} 
parallel electron-hole flat bands separated by their respective electron-like 
and hole-like dispersive bands. These parallel FBs turn out to be robust under unequal hopping strengths. 
We note that these flat bands at different energies in the absence of time-reversal symmetry can be leveraged to 
understand several physical phenomena such as quantum Hall phenomena at fractional filling, chiral spin 
liquid physics, topological transitions, etc~\cite{nagaosa-prb2000}. 

The rest of the paper is organized as follows. In Sec.~\ref{model}, we introduce and discuss the tight-binding
Hamiltonian of spinless particles moving in a two-dimensional lattice formed by fractal unit cells.
This is followed by Sec.~\ref{without-flux}, where we discuss different features of FBs and their neighboring
dispersive bands, in the absence of any external magnetic flux. Moreover, we demonstrate local density of states
and show distribution of wavefunction amplitudes at different lattice sites. 
In Sec.~\ref{with-flux}, we present the results for staggered and non-staggered magnetic flux.   
In Sec.~\ref{expt-realize}, we discuss the scope of possible experimental realization of our model, 
using photonic waveguide lattice structure. Finally, we conclude with a discussion on the usefulness of the proposed lattice model and 
possible future directions in Sec.~\ref{conclu}.

%%%%%%%%%%%%%%%%%%%%%%%%%%%%%%%%%%%%%%%%%%%%%%%%%%%%%%%%%%%%%%%%%%%%%%%%%%%%%%%%%%%%%%%%%%%%%%%%%%%%%%%%%%%%%%%%%%%%

\section{The model}
\label{model}
We propose a class of two-dimensional lattice models where segment of fractal 
geometries act as a constituent unit cell of the lattice structure. Fig.~\ref{fig:lattice} 
illustrates such a lattice structure where first generation ($\ell=1$) of a SPG fractal 
plays the role of the unit cell. Similar structures can be constructed using other 
higher generations of the SPG fractal geometry acting as the unit cell of the lattice. 
For simplicity, we focus on the $\ell=1$ generation of SPG with six atoms per unit cell. 
With this, the tight-binding Hamiltonian can be written in Wannier basis as,
%------------------------------------------------------------------------------------------------------------ 
\begin{align}
H=& \sum_{n,m}\Big[ \sum_{j=A}^{F} \epsilon_{j} c_{n,m,j}^{\dagger}c_{n,m,j} + \Big(t c_{n,m,A}^{\dagger}c_{n,m,D} \nonumber\\
&+ t c_{n,m,A}^{\dagger}c_{n,m,E} + t c_{n,m,B}^{\dagger}c_{n,m,D} + t c_{n,m,B}^{\dagger}c_{n,m,F} \nonumber\\ 
&+ t c_{n,m,C}^{\dagger}c_{n,m,F} + t c_{n,m,C}^{\dagger}c_{n,m,E} + t c_{n,m,D}^{\dagger}c_{n,m,E} \nonumber\\
&+ t c_{n,m,E}^{\dagger}c_{n,m,F} + t c_{n,m,F}^{\dagger}c_{n,m,D} + \gamma c_{n,m,A}^{\dagger}c_{n+1,m+1,B}\nonumber\\ 
&+ \gamma c_{n,m,A}^{\dagger}c_{n-1,m+1,C} + \gamma c_{n,m,C}^{\dagger}c_{n+2,m,B} + h. c. \Big) \Big],
\label{eq:hamil-wannier}
\end{align}
%------------------------------------------------------------------------------------------------------------
where $(n,m)$ stands for the cell index, and the letters $A, B, C, D, E, F$ identify  
different atomic sites within a cell as depicted in Fig.~\ref{fig:lattice}. $c_{n,m,j}^{\dagger}$ 
($c_{n,m,j}$) creates (annihilates) an electron in the $(n,m)$-th cell for atomic site 
$j$, $\epsilon_{j}$ is the on-site potential for a $j$-type of atomic site, $t$ is the 
intra-cell hopping amplitude, and $\gamma$ denotes the inter-cell hopping amplitude. 

%Discrete Fourier transformation,
%\begin{equation}
%c_{n,m,j} = \sum_{\bf k} c_{{\bf k},j}\: e^{i(k_{x}n+k_{y}m)}.
%\label{FT}
%\end{equation}
In $\bm{k}\equiv (k_x,k_y)$-space, Eq.~\eqref{eq:hamil-wannier} can be recast as,
\begin{equation}
H = \sum_{\bm{k}} \psi^{\dagger}_{\bm{k}} \mathcal{H}(\bm{k})\psi_{\bm{k}},
\label{eq:mon-hamil}
\end{equation}
where $
\psi^{\dagger}_{\bm{k}} \equiv
\left(\begin{matrix}
c^{\dagger}_{{\bm{k}},A}  &  c^{\dagger}_{{\bm{k}},B}  &  c^{\dagger}_{{\bm{k}},C}
&c^{\dagger}_{{\bm{k}},D}  &  c^{\dagger}_{{\bm{k}},E}  &  c^{\dagger}_{{\bm{k}},F}
\end{matrix}\right),
$ and $\mathcal{H}(\bm{k})$ is given by,
%------------------------------------------------------------------------------------------------------------
\begin{align}
\mathcal{H}({\bm k}) =
\left(\begin{matrix}
\mathcal{M}({\bm k})  &  \mathcal{T}\\
\mathcal{T}^{\dagger}  &  \mathcal{G} \\
\end{matrix}
\right),
\label{eq:ham0}
\end{align}
%------------------------------------------------------------------------------------------------------------
where $\mathcal{M}({\bm k})$ is a square matrix of dimension three for all $\ell$, and is given by,
%------------------------------------------------------------------------------------------------------------
\begin{align}
\mathcal{M}({\bm k}) =
\left(\begin{matrix}
\epsilon_{A} & \gamma e^{i(k_x+k_y)} & \gamma e^{-i(k_x - k_y)} \\
\gamma e^{-i(k_x+k_y)} & \epsilon_{B} & \gamma e^{-2ik_x} \\
\gamma e^{i(k_x - k_y)} & \gamma e^{2ik_x} & \epsilon_{C}  \\
\end{matrix}
\right).
\label{eq:mk}
\end{align}
%------------------------------------------------------------------------------------------------------------
Note that the off-diagonal matrix elements of $\mathcal{M}({\bm k})$ connect the nearest 
unit cells. In contrast, the matrix elements of $\mathcal{G}$ and $\mathcal{T}$ represent 
hopping within the same cell and hence independent of the momentum ${\bm k}$. 
For $\ell=1$, they are can be expressed in appropriate basis as,
%######################################################
\begin{figure}[ht]
\includegraphics[clip,width=0.49\columnwidth]{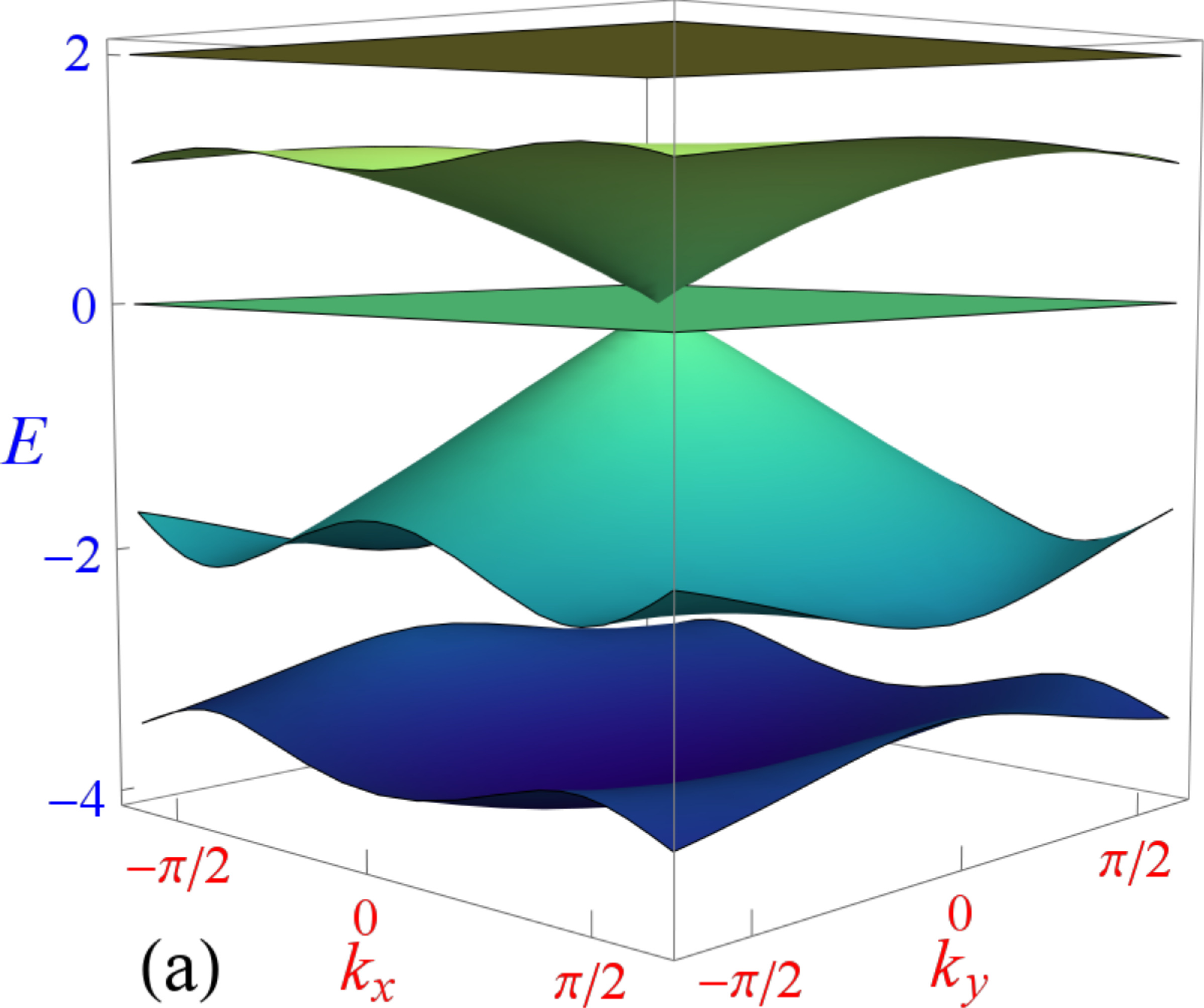}
\includegraphics[clip,width=0.49\columnwidth]{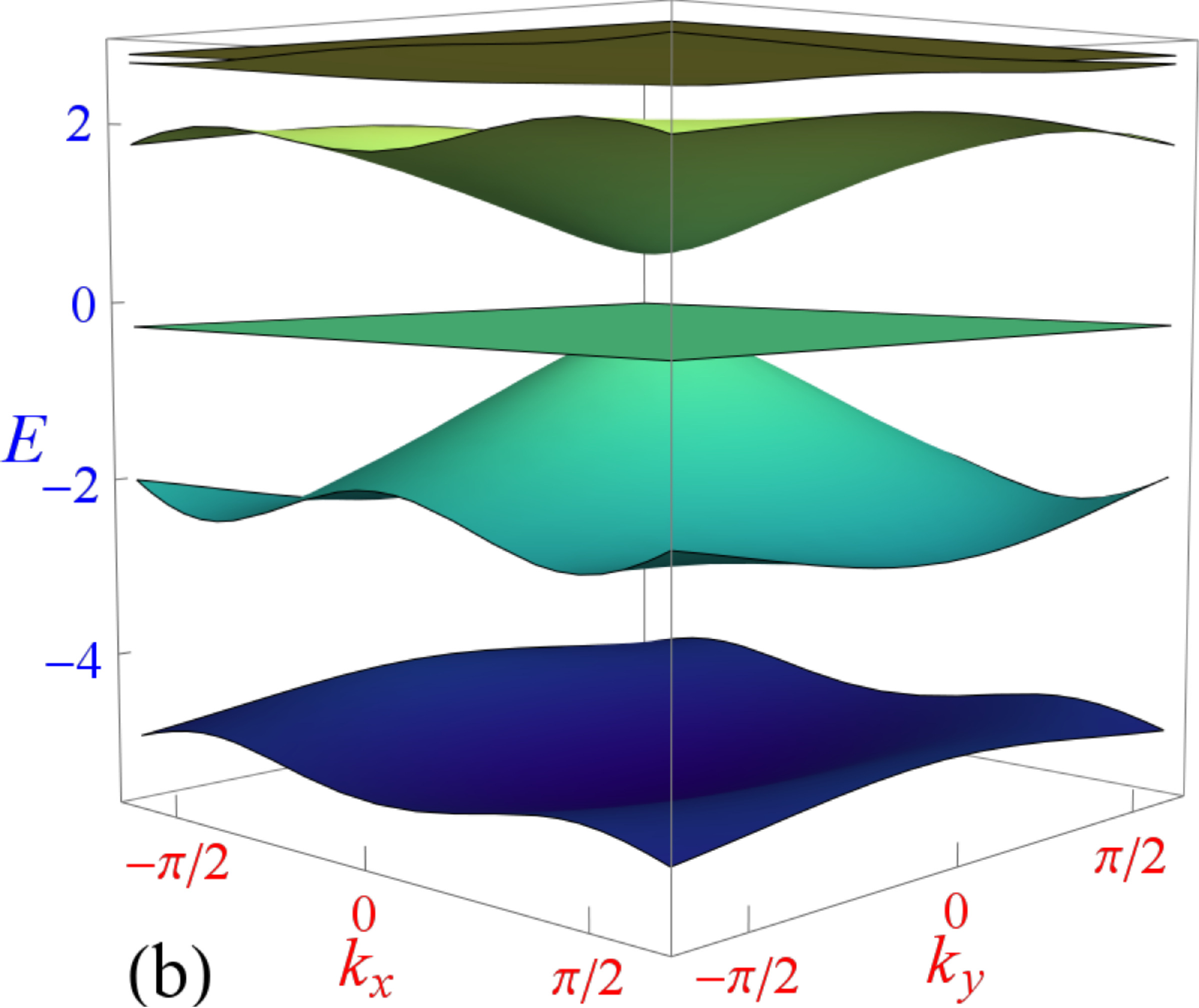}
\caption{Energy dispersions for $\ell=1$ generation SPG fractal network acting 
as the unit cell of the 2D lattice structure in Fig.~\ref{fig:lattice} with 
$\Phi_{\pm}=0$. The left panel corresponds to $t/\gamma = 1$ 
(with $t = -1$ and $\gamma = -1$), while the right panel is for 
$t/\gamma \ne 1$ (with $t = -1.5$ and $\gamma = -1$).}
\label{fig:no-flux-FB}
\end{figure}
%######################################################
%------------------------------------------------------------------------------------------------------------  
\begin{align}
\mathcal{G} =
\left(\begin{matrix}
\epsilon_{D}  &  t  &  t \\
t  &  \epsilon_{E}  &  t \\
t  &  t  &  \epsilon_{F}  \\
\end{matrix}
\right), \quad
\mathcal{T}=
\left(\begin{matrix}
t  &  t  &  0 \\
t  &  0  &  t \\
0  &  t  &  t \\
\end{matrix}
\right).
\end{align}
%------------------------------------------------------------------------------------------------------------

For generic $\ell$, $\mathcal{G}$ is a square matrix of dimension $(\mathcal{N}_{\ell} -3)$, while 
the dimension of $\mathcal{T} $ is found to be $(\mathcal{N}_{\ell} -3) \times 3$, 
where $\mathcal{N}_{\ell}$ is the number of atoms per unit cell. It is worth mentioning that Eq.~(\ref{eq:ham0})
is written in a basis which does not incorporate lattice symmetry (e.g., triangular symmetry). However, it can be 
shown that Hamiltonian in accordance with the lattice symmetry gives rise to similar spectrum as shown here.

To incorporate the effect of generic magnetic flux, we consider a flux $\Phi_{+}$ 
per ``up" triangular plaquette (light yellow shaded) in the SPG unit cell as shown 
in Fig.~\ref{fig:lattice}. Then the flux penetrating through the middle down triangle 
of the same unit cell turns out to be $-\Phi_{+}$. We further consider the flux penetrating through the 
``down" triangles (light blue shaded) connecting the two nearest SPG unit cells to 
be $\Phi_{-}$ as shown in Fig.~\ref{fig:lattice}. This fixes the flux through the 
hexagon to be equal to $-(2\Phi_{+} + \Phi_{-})$, when $\Phi_{\pm}$ are considered 
 {\it anticlockwise}. With this construction, the matrix elements $t$'s 
of $\mathcal{H}$ in Eq.~\eqref{eq:ham0} pick up a prefactor $e^{\pm i\Phi_{+}/3}$, while the 
$\gamma$'s pick up $e^{\pm i\Phi_{-}/3}$, where the sign of 
the phases is to be considered in accordance with the direction of the hopping.
%%%%%%%%%%%%%%%%%%%%%%%%%%%%%%%%%%%%%%%%%%%%%%%%%%%%%%%%%%%%%%%%%%%%%%%%%%%%%%%%%%%%%%%%%%%%%%%%%%%%%%%%%%%%%%%%%%%%

\section{Flat bands in zero flux}
\label{without-flux}
To analyze the energy spectrum for $\Phi_{\pm} = 0$, we diagonalize Eq.~\eqref{eq:ham0} 
and obtain six bands for $\ell=1$ generation SPG unit cell, containing multiple 
{\it completely flat} and dispersive bands. Depending on the relative strengths between 
the intra- and inter-cell hopping, these flat bands may be degenerate or non-degenerate. 
Moreover, the number of flat bands increases with the generation index, $\ell$.

\vspace{2mm}
{\it Equal intra and inter-cell hopping ---} For $t = \gamma$ and $\ell=1$, the flat bands energies are 
%---------------------------------------------------------------------------------
\begin{align}
E_{\rm FB} = -2t , -2t, 0.
\label{eq:fb-eng-l1-t-eq-g}
\end{align}
% E = 2, 2, 0
%---------------------------------------------------------------------------------
Fig.~\ref{fig:no-flux-FB}(a) shows that the doubly degenerate flat bands occur at the 
maximum of the spectrum and are {\it isolated} from rest of the other bands. 
This is in contrast to the commonly known ``frustrated hopping" model~\cite{bergman-prb2008} 
such as Kagome, where non-degenerate isolated flat-band can be obtained only in the 
presence of magnetic flux~\cite{green-prb2010}. The non-degenerate flat band at 
$E=0$ touches ``Dirac-point" formed by two electron-hole dispersive bands at 
${\bm k}=0$, and they all together form a spin-1 conical-type spectrum. These three 
bands at the center of the spectrum resembles with the spectrum of Kagome lattice with 
staggered flux~\cite{green-prb2010} or Lieb lattice with zero flux~\cite{manninen-pra2010, 
 vicencio-prl2015, mukherjee-prl2015, vicencio-njp2014, goldman-pra2011}. Note that, the ${\bm k}=0$ Dirac point 
 can be regarded as a special point where particle can have both zero effective 
mass (Dirac fermions) and infinite effective mass (FB state). Note also, the 
presence of this single Dirac point in the spectrum seems to violate fermion-doubling theorem. However, this 
is in consistent with the fact that fermion-doubling theorem can be avoided by introducing 
a flat band in the system~\cite{dagotto-plb1986}.  

\begin{figure}
\includegraphics[width=0.49\columnwidth]{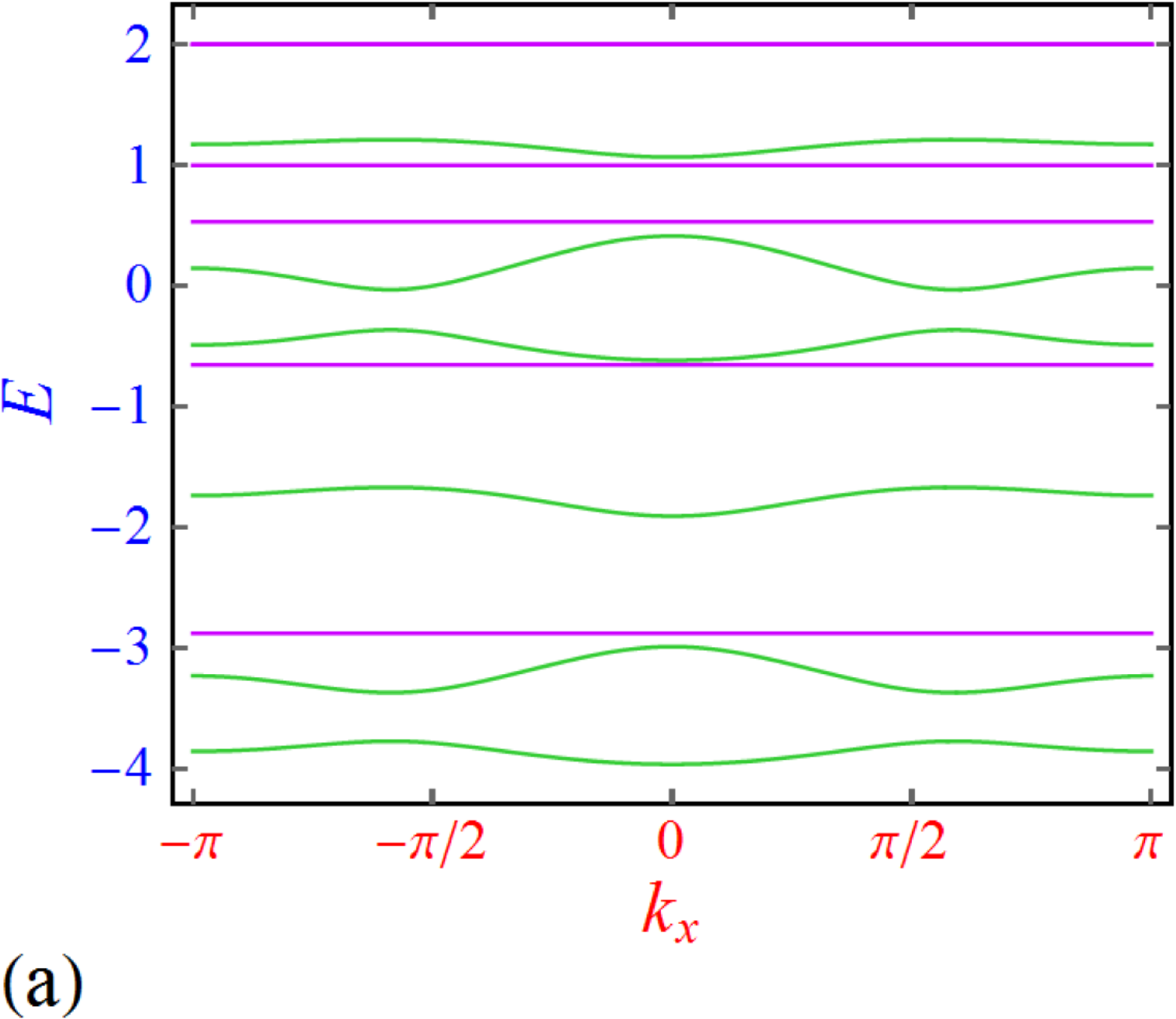}
\includegraphics[width=0.49\columnwidth]{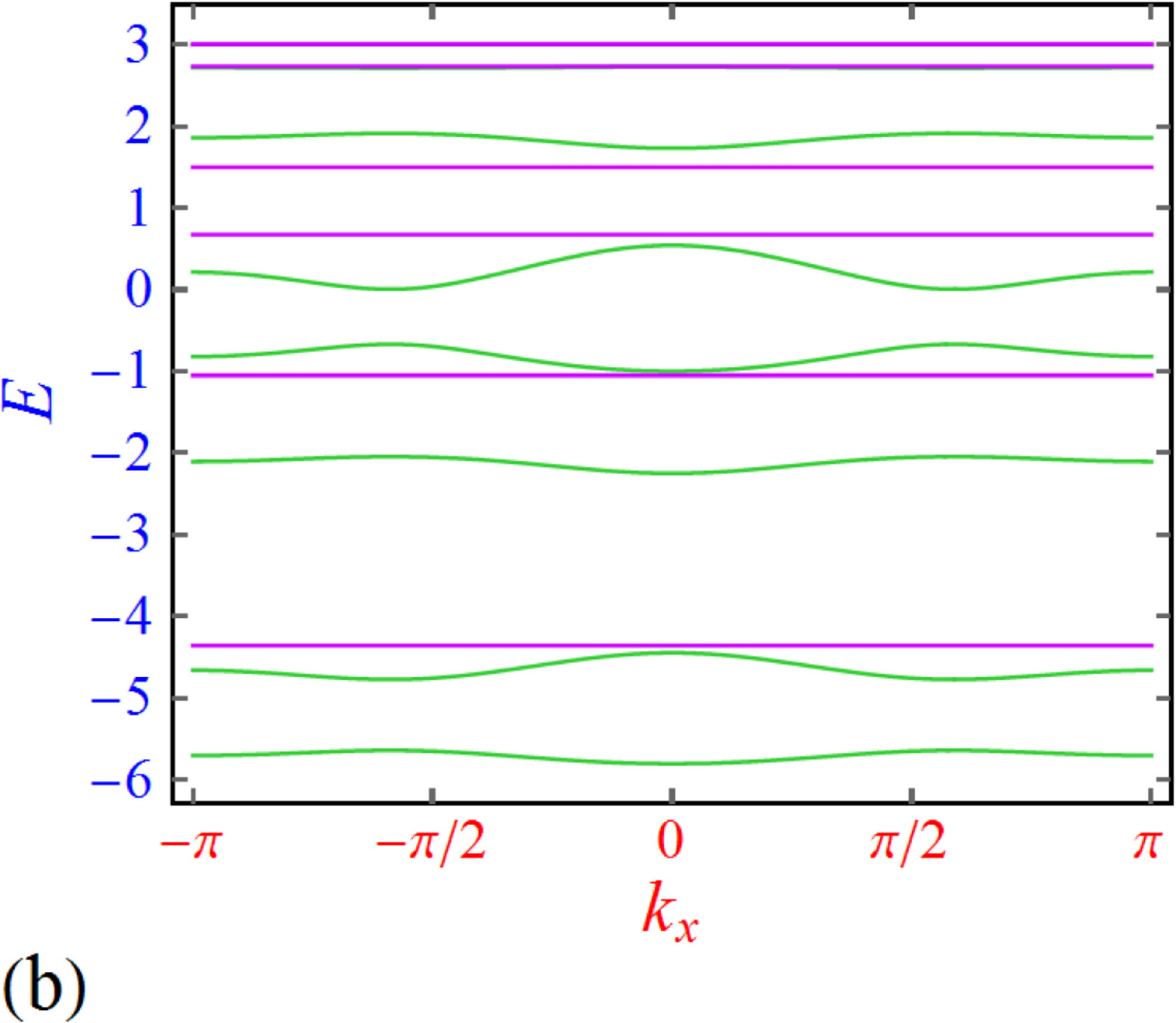}
\includegraphics[width=0.49\columnwidth]{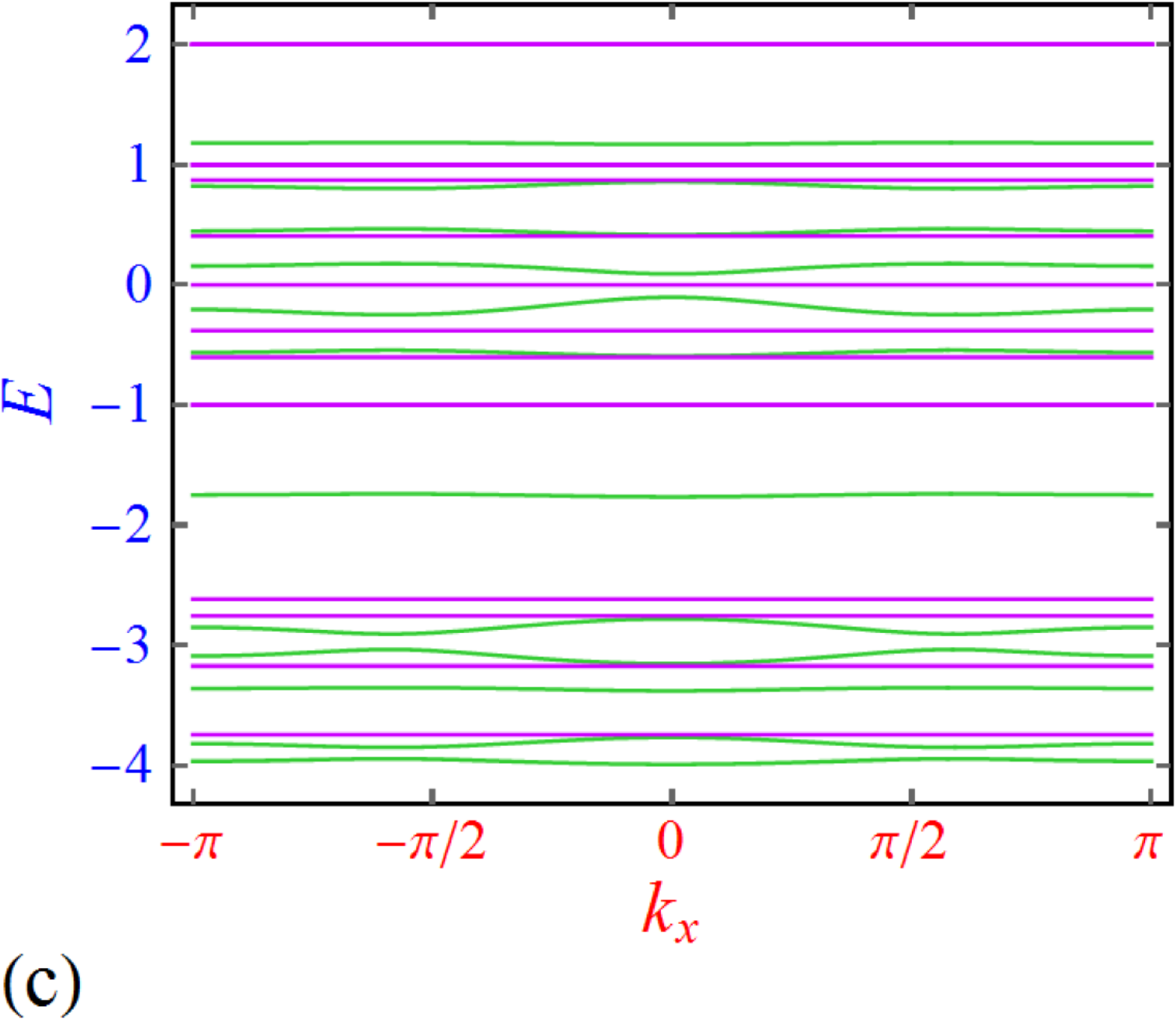}
\includegraphics[width=0.49\columnwidth]{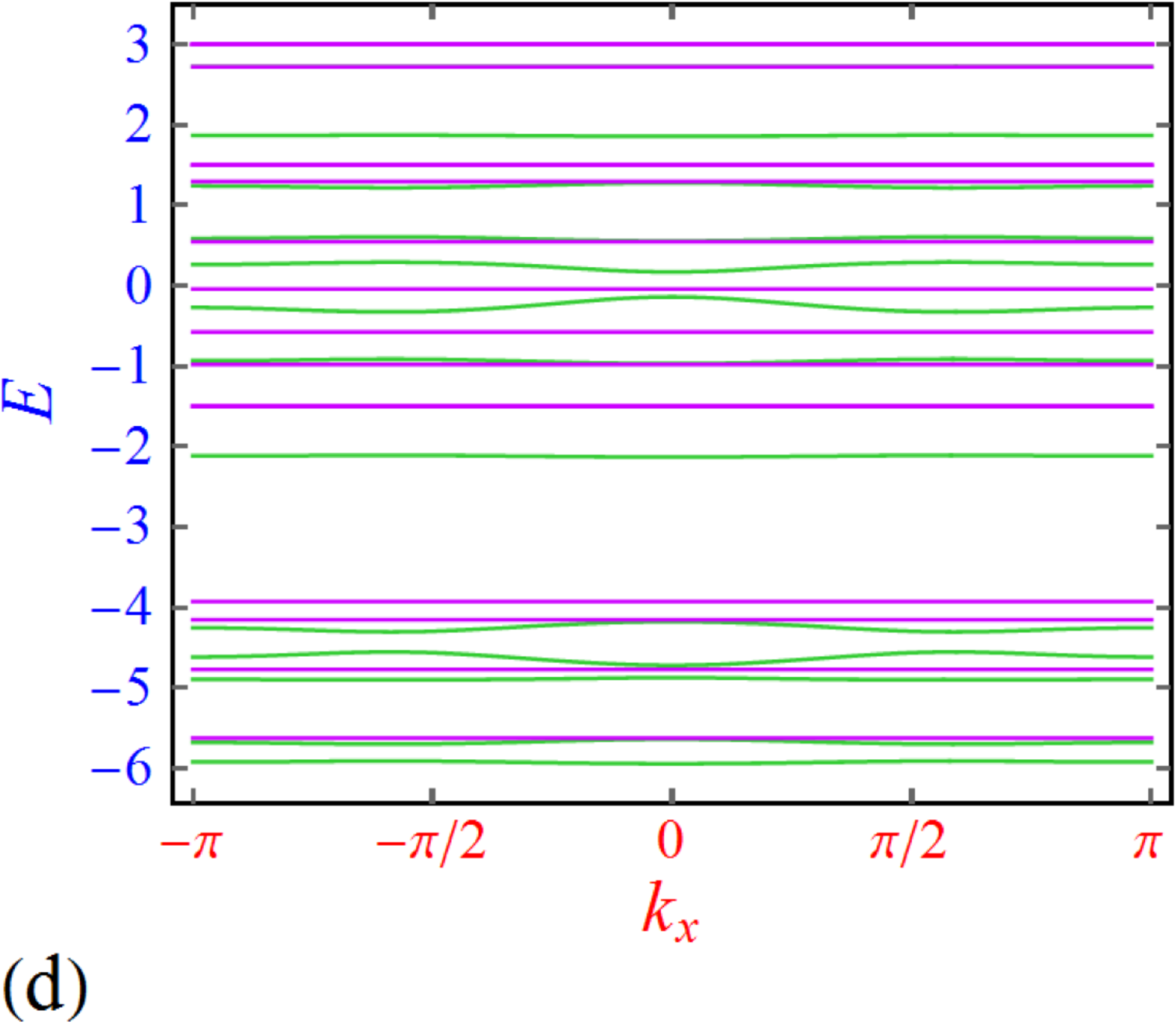}
\caption{$E$-$k$ diagram for (a-b) $\ell=2$, and (c-d) $\ell=3$ generation 
of SPG fractal unit cell. The {\it completely} flat bands are marked by the 
violet color flat lines, whereas dispersive or nearly flat bands are marked 
by green color. For the left panel the ratio between intra-cell and inter-cell 
hopping is taken to be $t/\gamma = 1$, whereas for the right panel it is 
$t/\gamma = 1.5$.}
\label{fig:2d-flat-bands}
\end{figure}

For $\ell = 2$ generation, the flat bands appear at the energies 
%---------------------------------------------------------------------------------
\begin{align}
E_{\rm FB} = -2t, -2t, -2t, -2t, -2t, -t, t f_{\alpha}(t), 
\label{eq:fb-eng-l2-t-eq-g}
\end{align}
% E = 2, 2, 2, 2, 2, 1, 0.53209, −0.65270, −2.87939
%---------------------------------------------------------------------------------
where $\alpha$ runs from $1$ to $3$, and $f(t)$ is some complicated function of $t$. 
Thus, we have five-fold degenerate FBs at $E = -2t$, and rest are non-degenerate with 
two electron-like and two hole-like FBs, as evident from Fig.~\ref{fig:2d-flat-bands}(a). 
We have exclusively checked energy spectrum for higher generations 
(see Fig.~\ref{fig:2d-flat-bands}(c)) and found that the number of FBs increases with 
the generation index $\ell$. In fact, we find a generalized formula to determine the 
number of FBs in a ($\ell + 1$)-th generation fractal. This is given by,
%--------------------------------------------------------------------------------------------------------
\begin{equation}
\mathcal{F}_{\ell+1} = \mathcal{N}_{\ell+1} - \mathcal{D}_{\ell+1} \quad \text{for } \ell \geq 0,
\label{FB-formula1}
\end{equation}
%--------------------------------------------------------------------------------------------------------
where $\mathcal{D}_{\ell+1}$ is the total number of dispersive bands (DBs) in a 
($\ell +1$)-th generation fractal, and is given by $
\mathcal{D}_{\ell+1} = 2\mathcal{D}_{\ell} $ for $\ell \geq 1$ with $\mathcal{D}_{1} = 3$. 
$\mathcal{N}_{\ell+1}$ is the total number of lattice sites in a ($\ell +1$)-th 
generation fractal unit cell, and is given by 
$\mathcal{N}_{\ell+1} = 3\mathcal{N}_{\ell} - 3$ for $\ell \geq 1$ with $\mathcal{N}_{1} = 6$. 
%######################################################
%######################################################

\vspace{2mm}
{\it Unequal intra and inter-cell hopping ---} In contrast to $t=\gamma$, the unequal 
intra and inter-cell hopping, \ie $t \neq \gamma$ reduces the number of flat bands 
and the degeneracy as well. 
For $\ell=1$, we obtain two FBs with energies (see Fig.~\ref{fig:no-flux-FB}(b))
%---------------------------------------------------------------------------------
\begin{equation}
\begin{aligned}
& E_{\rm FB} = \frac{1}{2} \left(-\gamma-t+\sqrt{(\gamma-t)^2+4t^2}\right)\\
& E_{\rm FB} = -\frac{1}{2} \left(\gamma+t+\sqrt{(\gamma-t)^2 + 4t^2}\right),
\end{aligned}
\label{fb-eng-l1-t-ne-g}
\end{equation}
% E = 2.77069, −0.27069
%---------------------------------------------------------------------------------
whereas for $\ell = 2$ generation, we obtain eight FBs (see Fig.~\ref{fig:2d-flat-bands}(b)) at energies 
\begin{align}
%---------------------------------------------------------------------------------
E_{\rm FB} = -2t, -2t, -2t, -t, f^{\prime}_{\alpha}(t,\gamma),
\end{align}
\label{fb-eng-l2-t-ne-g}
% E = 3, 3, 3, 1.5, 2.73209, 0.67477, −1.05065, −4.35620
%---------------------------------------------------------------------------------
where $\alpha$ runs from $1$ to $4$, and $f^{\prime}(t,\gamma)$ is some complicated 
function of $t$ and $\gamma$. Clearly, the number of FBs and the degeneracy are reduced. 
Also, unequal value of the hopping parameters lifts the degeneracy at the Dirac-point 
in conical-like spectrum since a finite gap appears between the FB and hole-like Dirac band 
(see Fig.~\ref{fig:no-flux-FB}(b)). However, the degeneracy between the FB and its 
neighboring electron-like dispersive band is retained at ${\bm k}=0$. This band 
touching turns out to be linear unlike the case in Kagome lattice. 
Note that, this band feature is reminiscent to the spectrum of 
artificial ice due to point-like dipole with fine-tuned external offset parameter 
related to sublattice~\cite{moller-prl2006}. Thus, for finite filling our lattice model
may reveal interesting spin-ice physics. Investigating higher generations (Fig.~\ref{fig:2d-flat-bands}) as before, 
we find a generalized formula for determining FBs for $t\ne\gamma$ as,
\begin{equation}
\mathcal{F}_{\ell+1} = \mathcal{N}_{\ell+1}-\mathcal{D}_{\ell+1} \quad \text{for } \ell \geq 0,
\label{eq:FB-formula2}
\end{equation}
where $\mathcal{D}_{\ell+1} = 2\mathcal{D}_{\ell} - 1$ for $\ell \geq 1$ with $\mathcal{D}_{1} = 4$, 
and $\mathcal{N}_{\ell+1}$ is same as before.

As discussed, the completely flat band states are generated by highly localized eigenstates. 
To corroborate this fact, we compute average density of states (ADOS) 
corresponding to the cases presented in Fig.~\ref{fig:no-flux-FB}(a) and~\ref{fig:no-flux-FB}(b). 
Using standard Green's function formalism, the ADOS is defined as,
%---------------------------------------------------------------------------------
\begin{equation}
\rho(E) = -\dfrac{1}{N\pi}\textrm{Im} \left[ \textrm{Tr}\; {\bm G}(E) \right],
\label{eq:ADOS}
\end{equation} 
%---------------------------------------------------------------------------------
where ${\bm G}(E) = \left[ (E+i\eta){\bm I} - H \right]^{-1}$ with $\eta$ being a small imaginary part added to 
the energy $E$, $N$ is total number of sites in the system, `$\textrm{Im}$' is the imaginary part, and `$\textrm{Tr}$' denotes 
the trace of the Green's function ${\bm G}$. 

%######################################################
\begin{figure}[ht]
\includegraphics[clip,width=0.48\columnwidth]{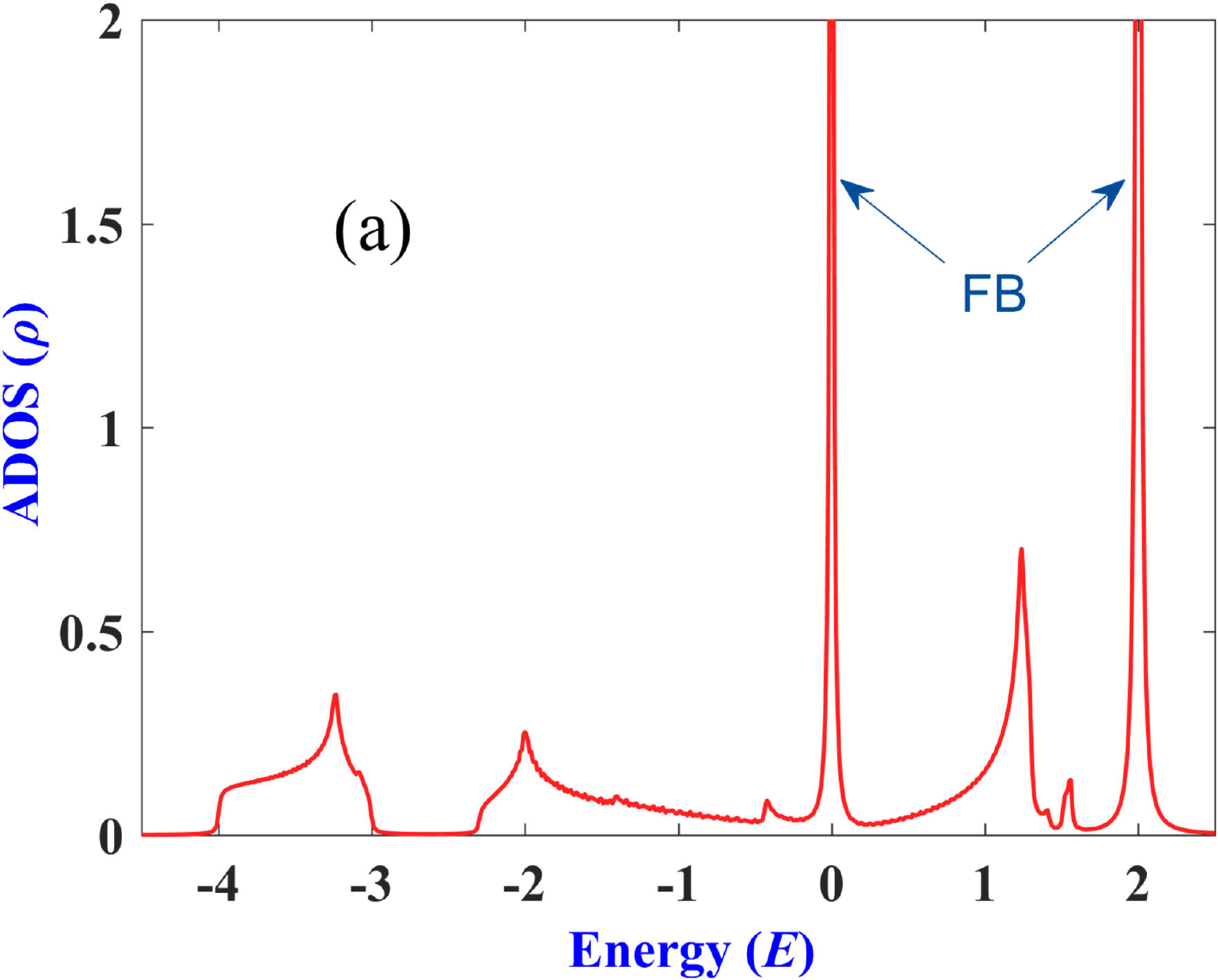}
\includegraphics[clip,width=0.48\columnwidth]{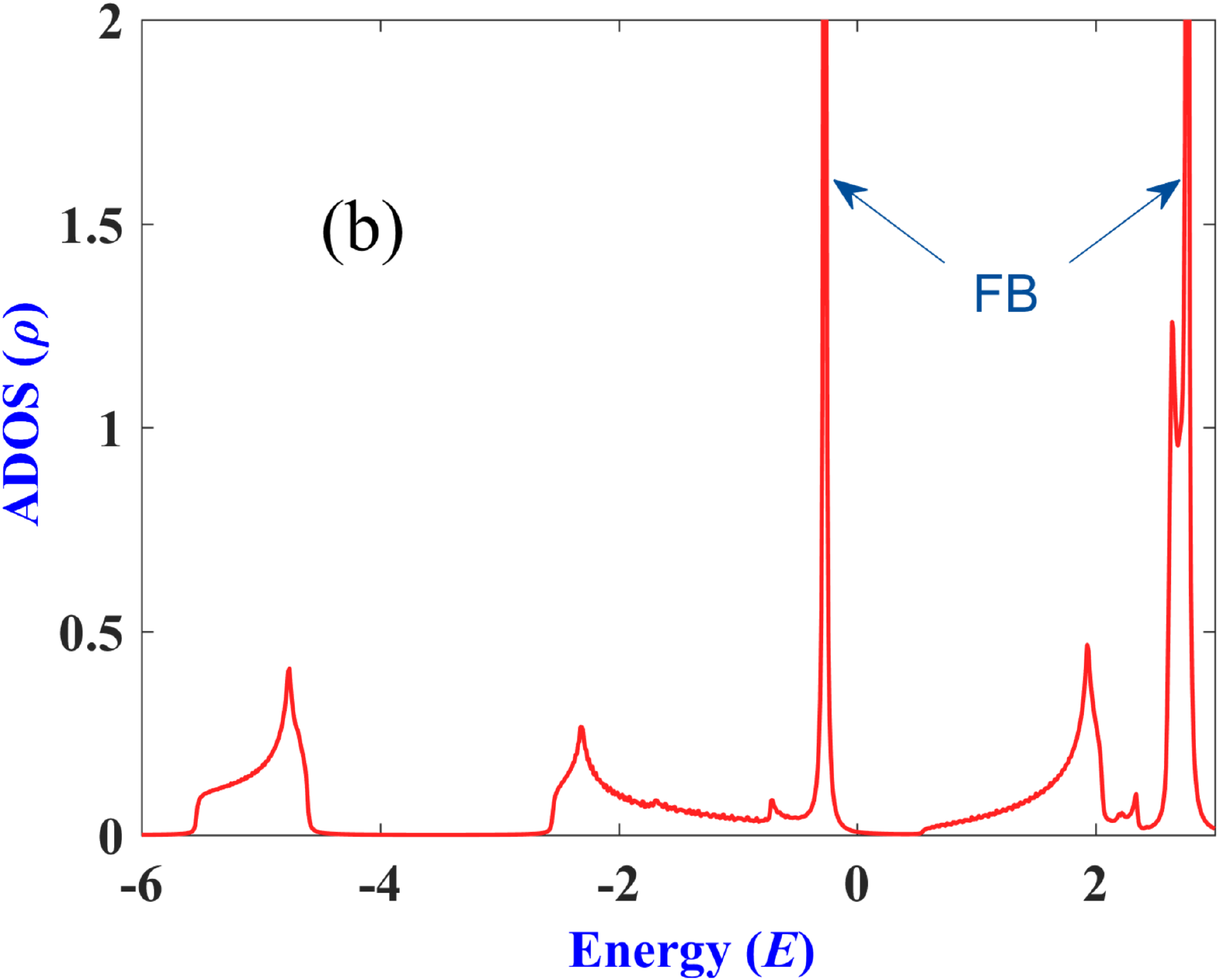}
\caption{Average density of states (ADOS) for $\ell = 1$ generation SPG unit cells repeated in 
$X$-$Y$ direction to form a 2D lattice structure with size  $L=(40 \times 40)$ as shown in Fig.~\ref{fig:lattice}. 
The left panel corresponds to $t/\gamma = 1$ 
(with $t = -1$ and $\gamma = -1$), while the right panel is for 
$t/\gamma \ne 1$ (with $t = -1.5$ and $\gamma = -1$). 
The states corresponding to the flat bands (FB) are indicated by the arrowheads.}
\label{fig:ADOS}
\end{figure}
%######################################################

Together with Eq.~(\ref{eq:ADOS}) and (\ref{eq:hamil-wannier}), we compute the ADOS for $\ell = 1$ 
generation SPG structure with system size $L=40$. 
Fig.~\ref{fig:ADOS}(a) and~\ref{fig:ADOS}(b) show very sharp localized states in the ADOS for field free case. 
The appearance of such localized states can be traced back to the presence of flat band states in 
the energy spectrum as shown in Fig.~\ref{fig:no-flux-FB}(a) and~\ref{fig:no-flux-FB}(b). 
We note that the ADOS for the cases with the magnetic field shows similar highly peaked localized states at 
corresponding flat band energies.  

We next turn to distribution of wave function amplitudes at different lattice sites,
which can be easily evaluated using the Schr\"odinger equation of the form,
%---------------------------------------------------------------------------------
\begin{equation}
\left( E  - \epsilon_{i} \right) \psi_{i}  = \sum_{j} \tau_{ij} \psi_{j},
\label{eq:difference-eq}
\end{equation}
%---------------------------------------------------------------------------------
where $\epsilon_{i}$ is the on-site potential at $i$-th site, $\tau_{ij}$ is the hopping amplitude between 
neighboring sites, and $\psi_{i}$ is the wave function amplitude  at $i$-th site. 
%######################################################
\begin{figure}[ht]
\includegraphics[clip,width=0.95\columnwidth]{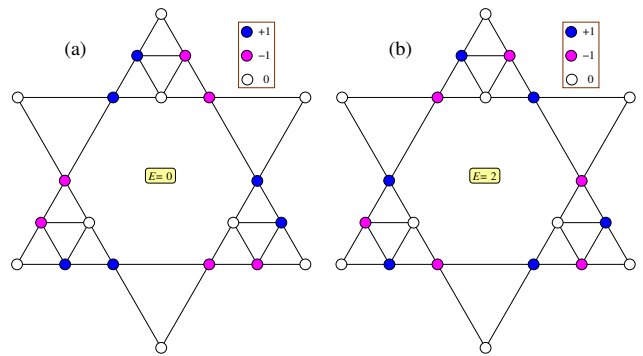}
\caption{Distribution of wave function amplitudes at different lattice sites 
corresponding to highly localized FB states with energies (a) $E=0$ and 
(b) $E=2$ as shown in Fig.~\ref{fig:ADOS}(a). The on-site potential 
for all the sites is set to zero, and the 
hopping parameters are $t = -1$ and $\gamma = -1$ respectively. 
The values of the wave function amplitudes on different lattice sites are 
$+1$ (blue/dark filled circles), $-1$ (magenta/light filled circles), and 
$0$ (empty circles).}
\label{fig:amplitude}
\end{figure}
%######################################################
Fig.~\ref{fig:amplitude}(a) and~\ref{fig:amplitude}(b) illustrates the wave function amplitude distribution 
corresponding to the FB states with energies $E=0$ and $E=2$ as shown in Fig.~\ref{fig:ADOS}(a). 
Evidently, the wave function corresponding to the FB states are 
localized over a finite number of lattice sites with non-zero amplitudes (marked by filled circles), and 
beyond those sites the amplitudes of the wavefunction turns  out to be zero (marked by empty circles). 
Using the same procedure, one can also figure out similar kind of wave function amplitude 
distribution pattern corresponding to the FB energies shown in Fig.~\ref{fig:ADOS}(b). 
%%%%%%%%%%%%%%%%%%%%%%%%%%%%%%%%%%%%%%%%%%%%%%%%%%%%%%%%%%%%%%%%%%%%%%%%%%%%%%%%%%%%%%%%%%%%%%%%%%%%%%%%%%%%%%%%%%%%

\section{Effect of magnetic flux on flat bands}
\label{with-flux}
The objective of this section is to investigate the fate of flat bands in the presence 
of staggered magnetic flux $\Phi_{\pm}$ (measured in units of the fundamental flux quantum 
$\Phi_{0} = h/e$) piercing through the triangular plaquettes in 
the fractal lattice geometry as shown in Fig.~\ref{fig:lattice}. It turns out that 
generic magnetic flux destroys FBs, leading to gapped dispersive bands. This destruction 
is independent of the relative strength between the hopping parameters $t$ and $\gamma$. However, for special 
values of $\Phi_{\pm}$, flat bands reappear and can be classified into three types.
%######################################################
\begin{figure}
\includegraphics[width=0.49\columnwidth]{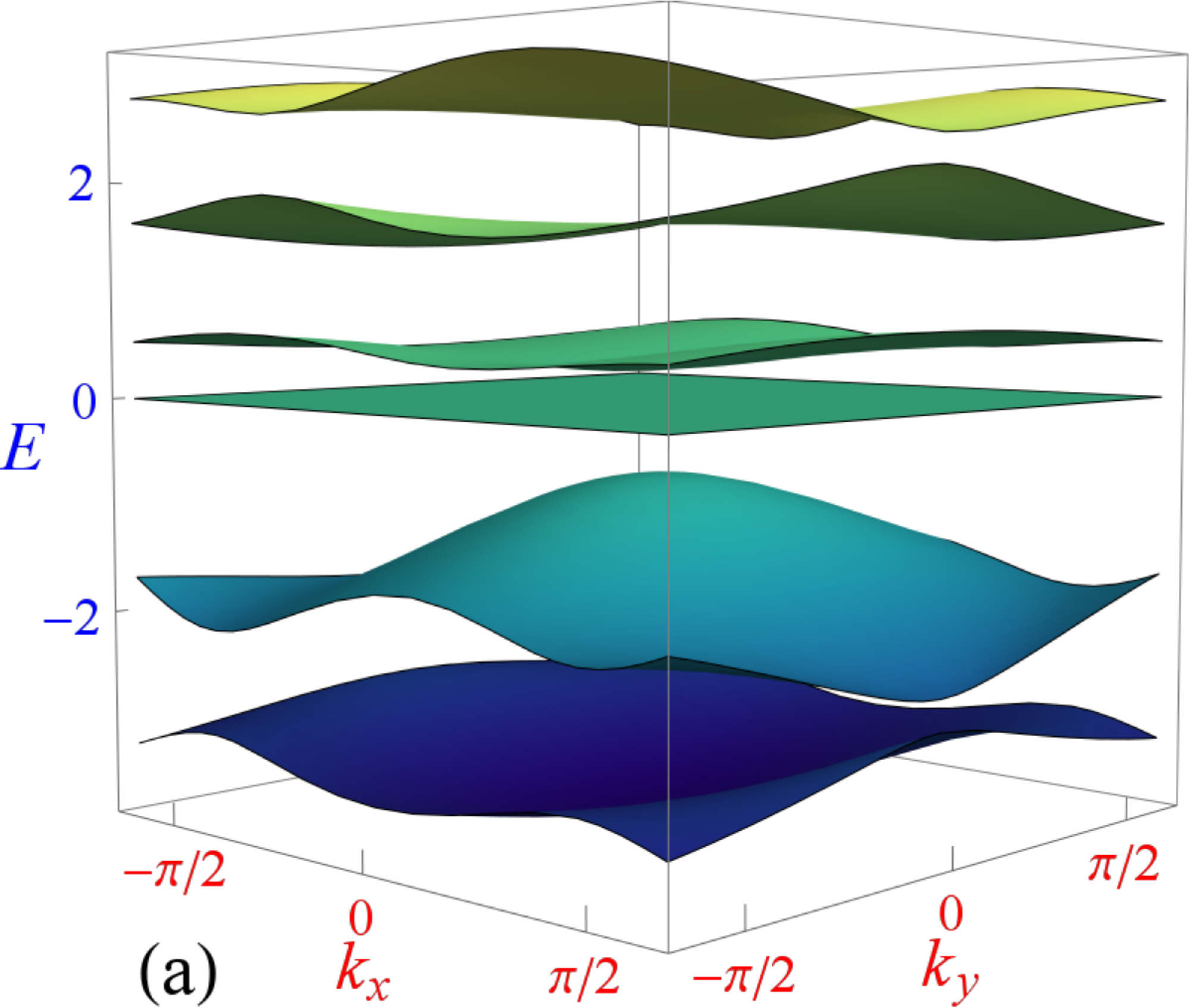}
\includegraphics[width=0.49\columnwidth]{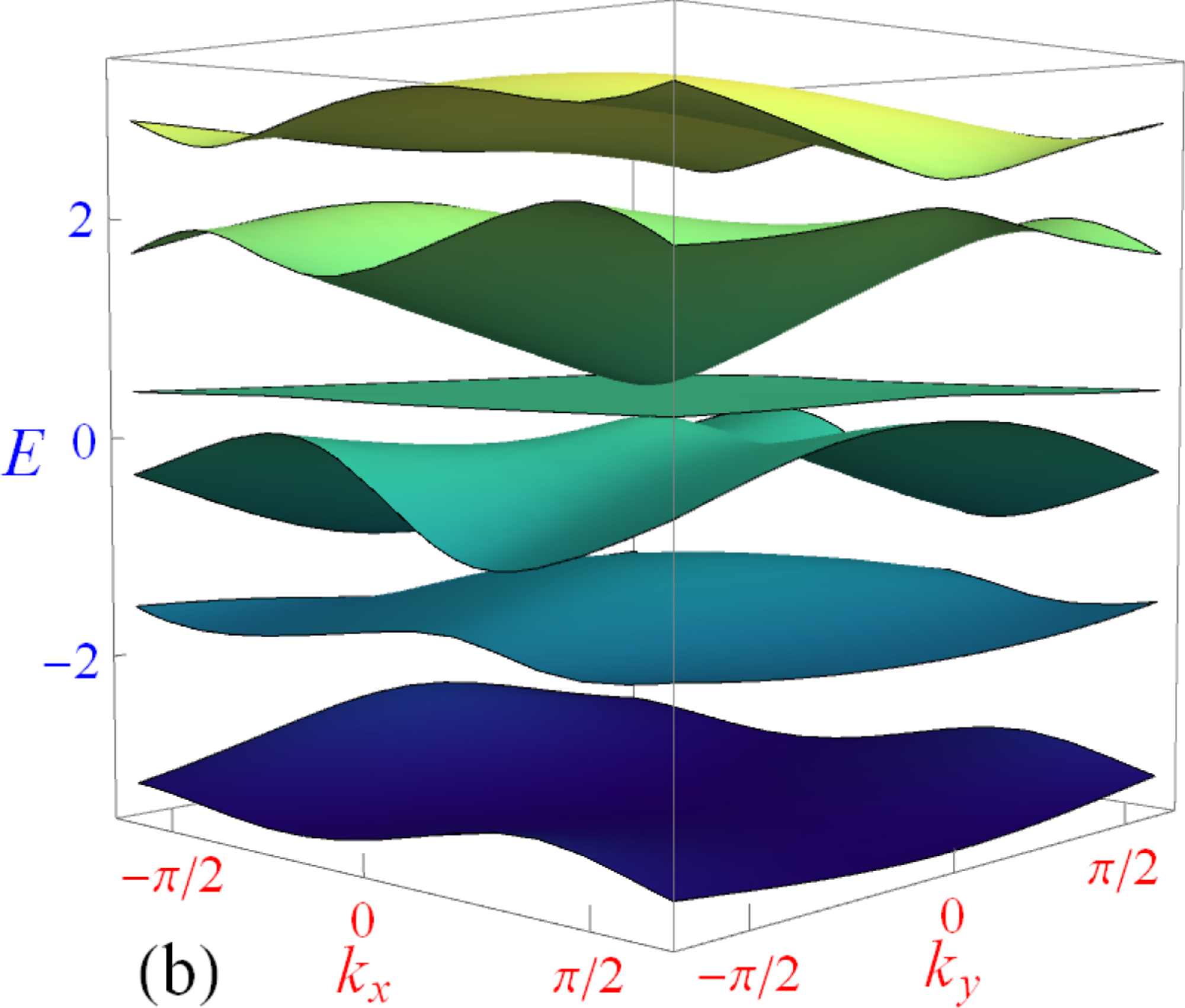}
\includegraphics[width=0.49\columnwidth]{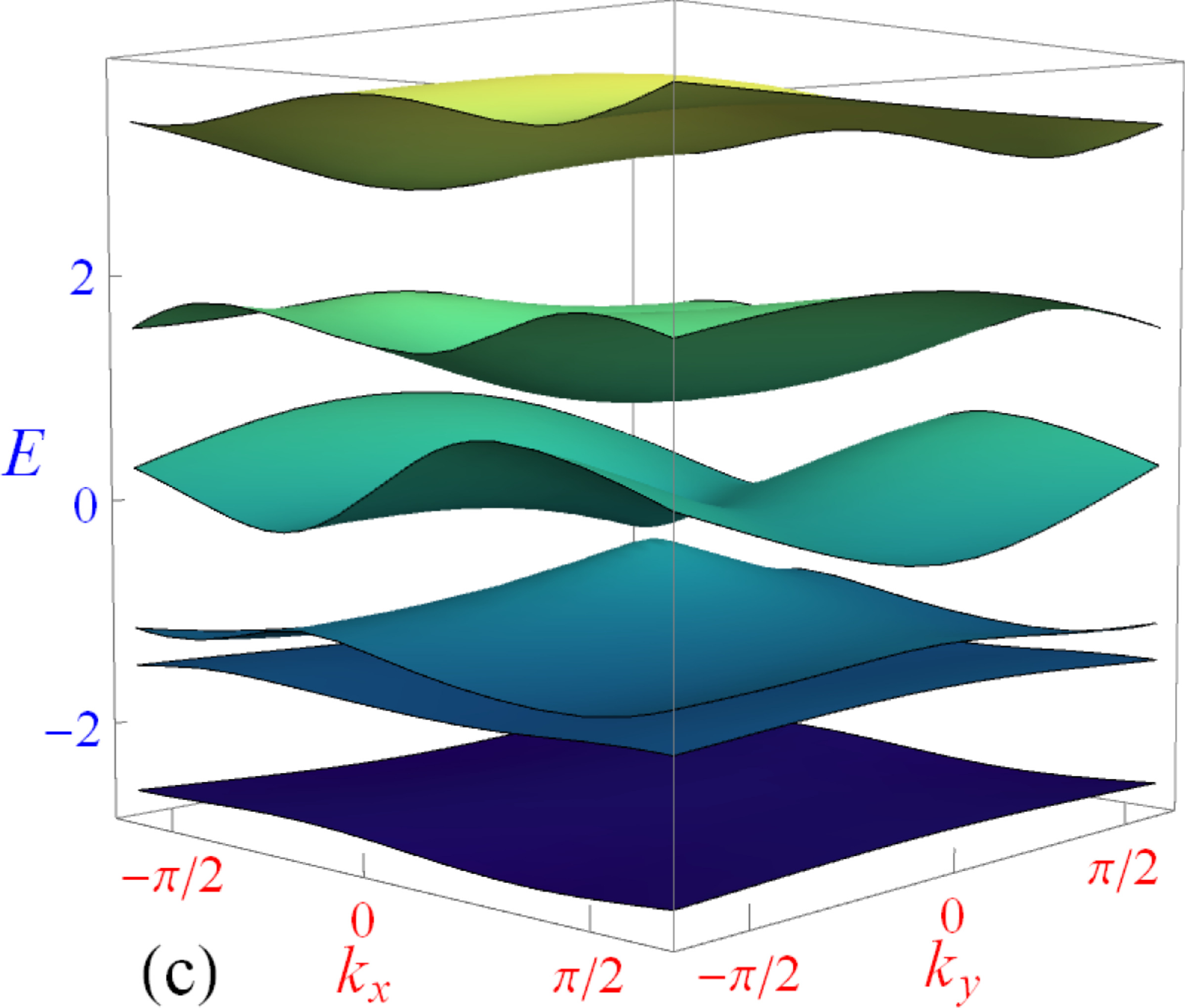}
\includegraphics[width=0.49\columnwidth]{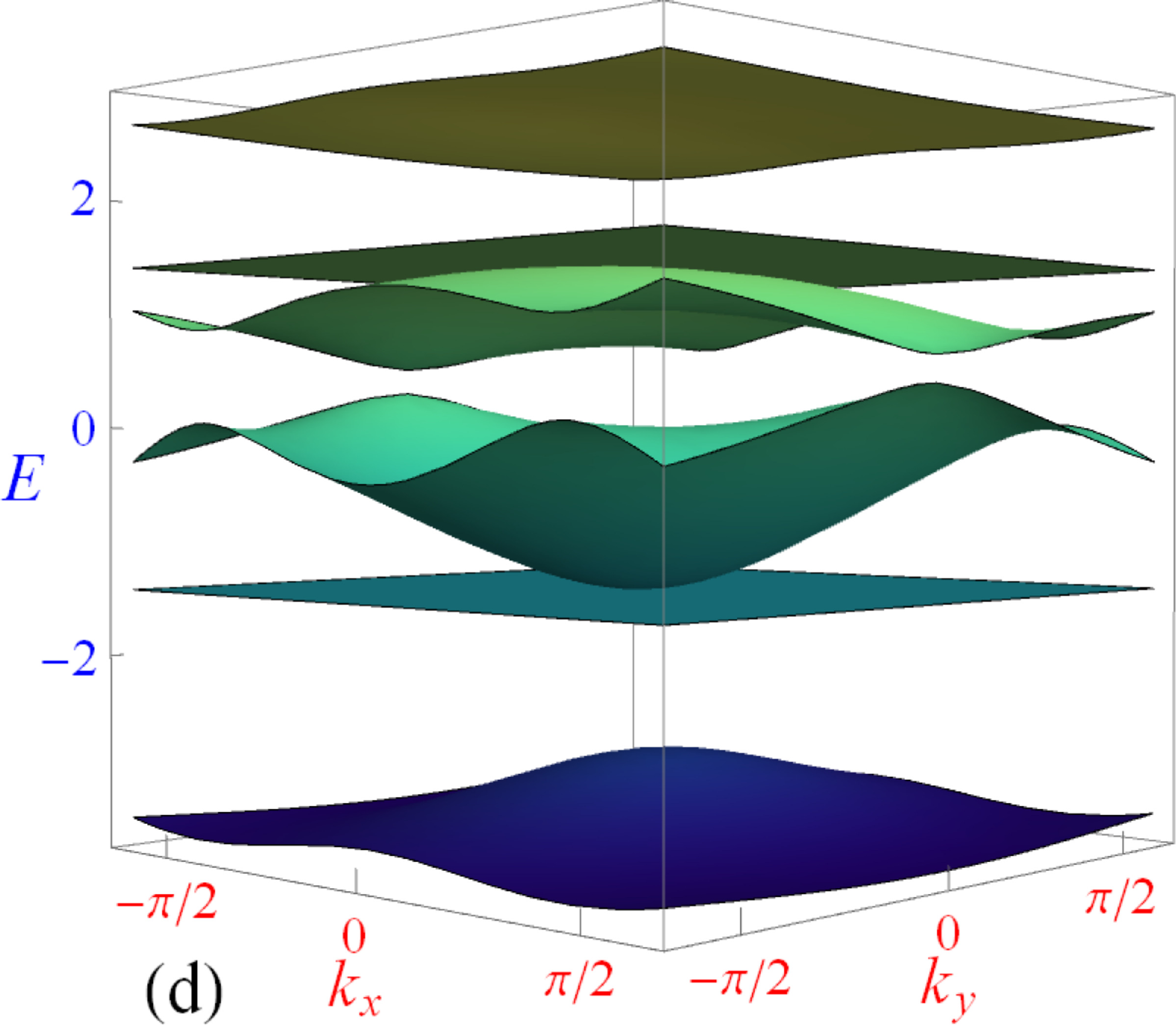}
\caption{Energy dispersions in presence of the staggered magnetic fluxes 
$\Phi_{\pm}$ for $\ell=1$ generation SPG fractal geometry. Different panels 
correspond to different combinations of fluxes $\Phi_{+}$ and $\Phi_{-}$, \viz 
(a) $\Phi_{+}=\Phi_{-}=\pi/3$, (b) $\Phi_{+}=\pi/3$ and $\Phi_{-}=-\pi$, 
(c) $\Phi_{+}=-\pi$ and $\Phi_{-}=\pi/3$, and (d) $\Phi_{+}=2\pi$ and $\Phi_{-}=\pi$. 
We have used $t/\gamma=1$.}
\label{fig:with-flux-FB}
\end{figure}
%###################################################### 

For $t=\gamma$ and $\Phi_{+} = \Phi_{-} = \pi/3$, we obtain a single isolated flat band 
at $E_{\rm FB}=0$, separating three hole-like and two electron-like 
dispersive bands, as shown in Fig.~\ref{fig:with-flux-FB}(a). The two electron-like 
dispersive bands resemble the spectrum of gapped graphene in the absence of time-reversal 
symmetry~\cite{haldane-prl1988}. Since this isolated flat band is protected by the gap, 
the degenerate states can be used to form correlated states when interactions are added. 
Note that, the isolated FB acquires a small curvature as intra-cell and inter-cell 
hopping differs from each other. 

In contrast, for $\Phi_{+} = \pm\pi/3$ and $\Phi_{-} = \mp\pi$, we obtain a single 
gapless flat band with energy $E_{\rm FB}>0$. Moreover, this FB touches its nearest 
dispersive band at a single point in the Brillouin zone (see Fig.~\ref{fig:with-flux-FB}(b)), 
and the band touching turns out to be linear similar to the band touching discussed in the 
preceding section. However, the band touching can be removed by unequal intra-cell and 
inter-cell hopping, keeping the FB {\it nearly} flat. For a reverse combination of 
$\Phi_{+}$ and $\Phi_{-}$, we do not see any gapless flat bands, rather two nearly 
FBs appear at the end of the band spectrum as shown in Fig.~\ref{fig:with-flux-FB}(c).

At $\Phi_{+} =  \Phi_{-} = \pm \pi$, the spectrum is particle-hole conjugate of the case 
at $\Phi_{\pm}=0$. With this combination of fluxes, the total flux through the individual 
unit cell and the down triangle connecting nearest unit cells turns out to be 
$(2\Phi_{+}-\Phi_{-})=\pm 3\pi$ or $\pm\pi$. The first condition is exactly similar to 
the condition for staggered fluxes in ``up'' and ``down'' triangles of Kagome 
lattice~\cite{green-prb2010} as a manifestation of time-reversal symmetry, while the 
latter is typical time-reversal symmetric non-staggered case of $n\pi$ flux through 
each plaquette, where $n$ is integer.      
 
In addition to the above combinations of $\Phi_{\pm}$'s, we obtain a particular combination 
of fluxes such that $(\Phi_{+}+\Phi_{-})=\pi$ (with $\Phi_{+}=0\ ({\rm or}\ \pi)$ and 
$\Phi_{-}=\pi\ ({\rm or}\ 0)$), or $3\pi$ (with $\Phi_{+}=2\pi\ ({\rm or}\ \pi)$ and 
$\Phi_{-}=\pi\ ({\rm or}\ 2\pi)$), for which two flat bands are separated 
by two dispersive bands. Fig.~\ref{fig:with-flux-FB}(d) illustrates 
this specific case, where we see both the electron-like and hole-like flat bands 
touch their nearest dispersive bands at a single point in the first BZ. Interestingly, the 
nature of band touching is quite different for these two FBs. While one FB touches quadratically 
to its nearest dispersive band, the other one touches the dispersive band linearly. Moreover, 
both of these band touching are protected from any perturbation in the Hamiltonian. Notice that, 
in this particular case, there  exists an isolated nearly flat band at the maximum or minimum of 
the spectrum. 

It is worth pointing out that there are some other combinations of fluxes for 
which a flat band and its nearest dispersive band overlap, which we do not elaborate here in 
order to not deviate from the main interesting features discussed in the preceding sections. 
We would also like to mention that similar distinct band features can be obtained
for higher generation fractal structures in presence of the external staggered flux. 
%######################################################
\begin{figure}[ht]
\includegraphics[clip,width=0.8\columnwidth]{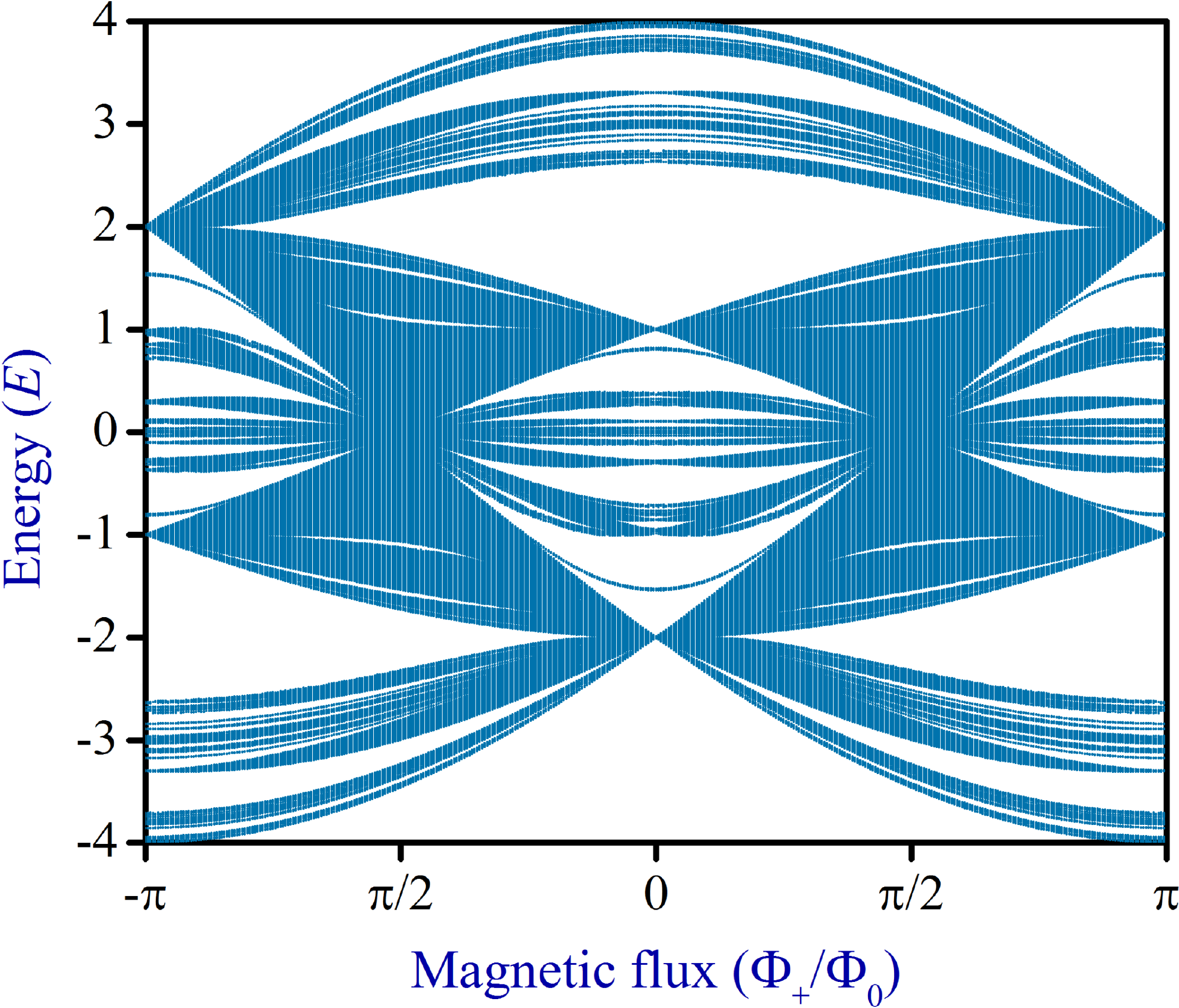}
\caption{Energy eigenvalue spectrum of a SPG fractal geometry in the 
thermodynamic limit as a function of the magnetic flux $\Phi_{+}$ 
(measured in units of the fundamental flux quantum $\Phi_{0} = h/e$). We 
have set the on-site potentials at all sites to be equal to zero and 
the hopping parameter $t$ is taken equal to 1.}
\label{fig:e-phi}
\end{figure}
%######################################################

Before ending this section, we briefly discuss the real space energy  
spectrum of a {\it single} unit cell of the SPG fractal structure in the thermodynamic limit as a function of magnetic flux. 
We consider each smallest ``up" triangular 
plaquettes in the unit cell carries flux $\Phi_{+}$. Then following Ref.~\cite{pal-prb2012}, we compute the energy 
eigenvalues as a function of $\Phi_{+}$. Fig.~\ref{fig:e-phi} illustrates the formation of bands 
and gaps in the spectrum. Clearly, the thinning of bands at $\Phi_{+}=0$ 
and its conjugate case, \ie $\Phi_{+}=\pi$ is attributed to the presence of FBs. 
%%%%%%%%%%%%%%%%%%%%%%%%%%%%%%%%%%%%%%%%%%%%%%%%%%%%%%%%%%%%%%%%%%%%%%%%%%%%%%%%%%%%%%%%%%%%%%%%%%%%%%%%%%%%%%%%%%%%

\section{Possible experimental realization}
\label{expt-realize}
\begin{figure}[ht]
\includegraphics[clip,width=\columnwidth]{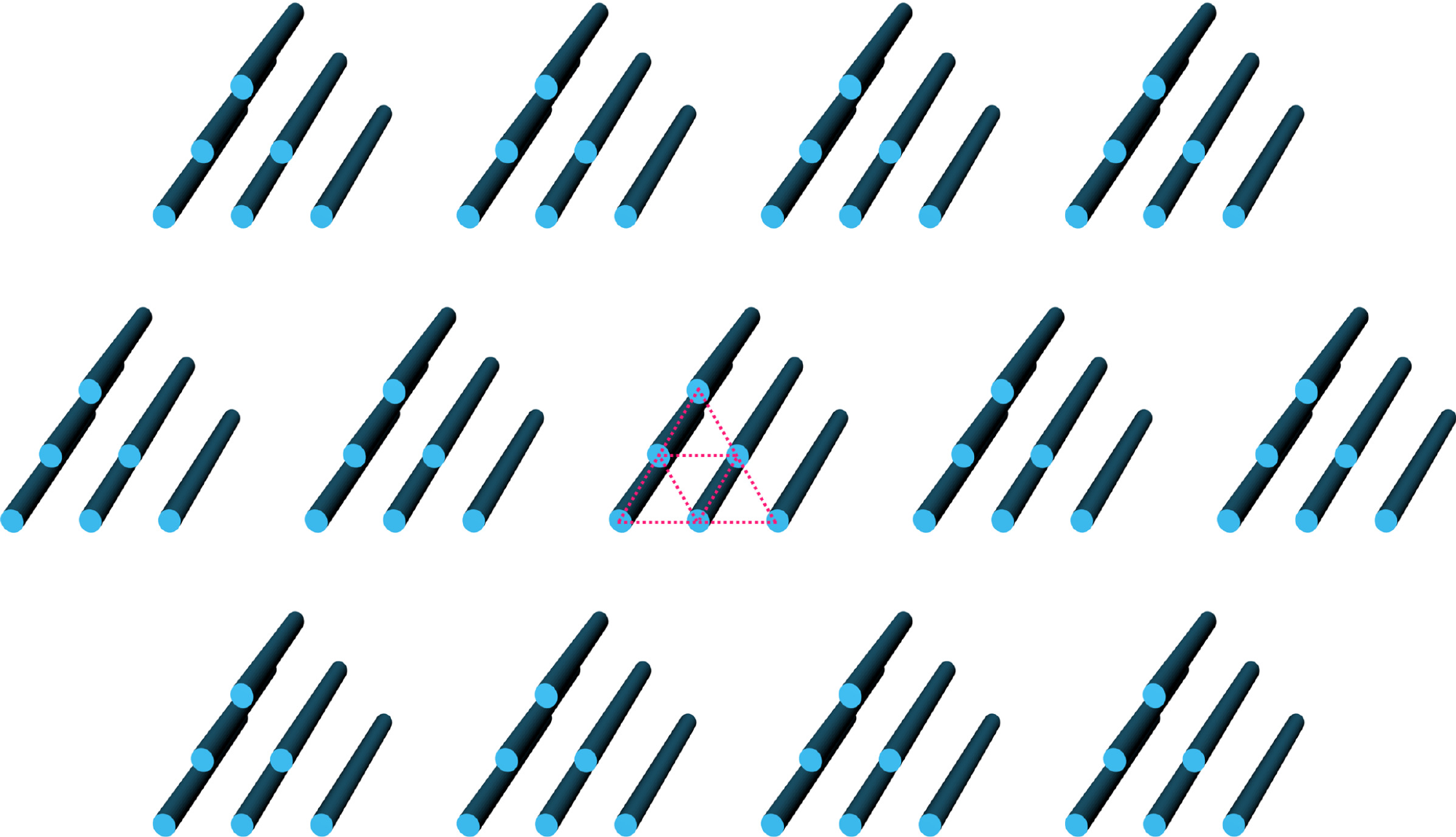}
\caption{Schematic diagram of a possible proposed photonic waveguide 
structure corresponding to the lattice structure shown in Fig.~\ref{fig:lattice}.}
\label{photonic-structure}
\end{figure}

Finally, we discuss the scope for possible experimental realization of our proposed 
complex lattice structure and related phenomena. In the spirit of recent experiments 
by Mukherjee \etal~\cite{mukherjee-ol2015, mukherjee-prl2015} and also by 
others~\cite{vicencio-prl2015,xia-ol2016,zong-oe2016}, photonic lattices formed by 
laser-induced single-mode waveguides can be used to study different FB properties 
discussed here, in the absence of undesired excitations such as phonons.  
In such photonic structure, the atomic sites in our proposed 2D fractal lattice geometry 
can be replaced by single-mode optical waveguides, which can be controlled by femtosecond laser-writing 
technique as well as the optical induction technique. This may allow one for direct observations 
of diffractionless FB states. A schematic diagram of the possible corresponding 
photonic waveguide structure of our proposed model (for $\ell=1$ generation SPG 
geometry acting as the unit cell of the lattice) is illustrated in 
Fig.~\ref{photonic-structure}. The intra- and inter-cell hopping parameters can be 
controlled by the refractive index of the lattice structure. Moreover, by 
modulating longitudinal propagation constants~\cite{longhi-ol2014, longhi-ol2013} 
of the waveguides, a synthetic gauge field can be generated to study the effect of 
magnetic the flux in our proposed lattice geometry. In addition to the photonic lattices, the unprecedented 
controllability of cold atoms in optical lattices may help to engineer such artificial 
complex structures in experiments, and study the desired properties in a very 
controllable and clean environment. Formation of similar complex fractal flat band 
waveguide networks has been proposed theoretically recently~\cite{atanu-pra2016}. 
%######################################################
%######################################################
%%%%%%%%%%%%%%%%%%%%%%%%%%%%%%%%%%%%%%%%%%%%%%%%%%%%%%%%%%%%%%%%%%%%%%%%%%%%%%%%%%%%%%%%%%%%%%%%%%%%%%%%%%%%%%%%%%%%

\section{Concluding remark and outlook}
\label{conclu}
In conclusion, we have studied flat bands in a complex 2D lattice structure formed by different generations of a SPG fractal geometry as 
the building unit cell of the lattice. We show that such complex structure gives rise to band spectrum with notable features,
such as appearance of isolated gapped FB states, or the presence 
of gapless single or multiple electron-like or hole-like FB states touching their nearest dispersive 
bands, or the formation of ``Dirac-cone" with a FB state sandwiched in-between them, resembling 
the spin-1 conical-type energy spectrum. We furthermore show that the different combinations of 
staggered or non-staggered flux may tune or detune the FB states, appearing in such a 
complex fractal-like lattice structure. For a particular flux, we  show that it is possible to separate a FB state 
from the other dispersive bands by a gap due to the fact that time-reversal symmetry is broken. 
In such situation, the FB state can be thought of as a critical point (with zero band curvature) that separates electron-like 
and hole-like bands with opposite band curvatures, and such transition may lead to an 
anomalous Hall effect~\cite{green-prb2010, nagaosa-prb2000}. We also show that in 
some cases, FB is degenerate with one or more other bands at a single point. Such 
band touching is accidental and can be removed by means of some perturbations.

In addition to these interesting band features, we establish a generic formula 
to determine the number of FB states as a function of  the fractal generation index, $\ell$. 
Finally, we address the possibility of realizing our model in a photonic waveguide network which 
can be fabricated by  using single mode optical 
waveguides in highly controllable environment using femtosecond laser 
pulses~\cite{vicencio-prl2015, mukherjee-prl2015}. 

The fractal FB model proposed by us 
has gone beyond the paradigm of regular flat band lattice geometries such as Lieb lattice,
hexagonal lattice, Kagome lattice, etc., and can be a new prospective lattice model to explore 
different physical phenomena. For example, topological protection of band touching in the 
majority of geometrically frustrated systems are known through a careful counting of linearly 
independent localized states~\cite{bergman-prb2008}. Since our proposed model exhibits both 
linear and quadratic band touching in a single lattice setting for a particular configuration of the 
parameters, topological protection in these settings may differ significantly from the typical 
frustrated lattice systems as discussed in Ref.~\cite{bergman-prb2008}. Note that such 
study is required to understand the physics of interacting bosons and fermions in flat bands 
with various filling for short-range interactions of arbitrary strength. 

On a related note, our 
proposed model with $\ell=1$ resembles to the lattice structure of a new class of magnetic 
materials, namely triangular Kagome lattices. This lattice structure is formed by extra set 
of triangles inside the Kagome triangles.  While solving Hubbard model in this ``triangle-in-triangle" 
lattice structure, several new phases such as plaquette insulator, Kondo metal can be obtained 
while asymmetry is introduced in the system~\cite{chen-prl2012}. In view of that, we believe 
our proposed model may give rise to even richer phases as a function of interaction, 
temperature and asymmetry. Overall, our findings provide a concrete framework for future studies, addressing the nature of  many-body ground state
in the presence of repulsive interactions and at fractional filling. Moreover, it may grow 
interests to explore the possibility of having FB states in  lattice structures formed by other 
interesting fractal geometries such as Vicsek and Koch fractal.

%%%%%%%%%%%%%%%%%%%%%%%%%%%%%%%%%%%%%%%%%%%%%%%%%%%%%%%%%%%%%%%%%%%%%%%%%%%%%%%%%%%%%%%%%%%%%%%%%%%%%%%%%%%%%%%%%%%%
%\vspace{5mm} 
%%%%%%%%%%%%%%%%%%%%%%%%%%%%%%%%%%%%%%%%%%%%%%%%%%%%%%%%%%%%%%%%%
\begin{acknowledgments}
BP would like to acknowledge  Anne E. B. Nielsen and Arunava Chakrabarti for 
useful discussions on related topics. We thank the referees for constructive comments. 
\end{acknowledgments}
%%%%%%%%%%%%%%%%%%%%%%%%%%%%%%%%%%%%%%%%%%%%%%%%%%%%%%%%%%%%%%%%%

\end{document}